\documentclass[lettersize,journal]{IEEEtran}
\usepackage{amsmath,amsfonts}
\usepackage{array}
\usepackage[caption=false,font=normalsize,labelfont=sf,textfont=sf]{subfig}
\usepackage{textcomp}
\usepackage{stfloats}
\usepackage{url}
\usepackage{verbatim}
\usepackage{graphicx}
\usepackage{cite}
\usepackage{amssymb}
\usepackage{pgffor}
\usepackage{caption}
\usepackage{amsthm}
\usepackage{color}
\usepackage{cite}
\usepackage{amsthm,amsmath,amssymb}
\usepackage{graphicx} 
\usepackage{epstopdf}
\usepackage[linesnumbered, ruled]{algorithm2e}
\usepackage{amssymb}
\SetKwRepeat{Do}{do}{while}%
\usepackage{bm}

\usepackage{mathrsfs}

\begin{document}
	
	\title{RIS-assisted High-Speed Railway Integrated Sensing and Communication System}
	
	\author{Panpan Li, Yong Niu,~\IEEEmembership{Member,~IEEE}, Hao Wu,~\IEEEmembership{Member,~IEEE}, Zhu Han,~\IEEEmembership{Fellow,~IEEE}, Guiqi Sun,
		
		 Ning Wang,~\IEEEmembership{Member,~IEEE}, and Zhangdui Zhong,~\IEEEmembership{Fellow,~IEEE,} Bo Ai,~\IEEEmembership{Fellow,~IEEE}
		
		\thanks{
			Copyright (c) 2015 IEEE. Personal use of this material is permitted. However, permission to use this material for any other purposes must be obtained from the IEEE by sending a request to pubs-permissions@ieee.org. This work was in part by the Fundamental Research Funds for the Central Universities 2023JBMC030; in part by the Fundamental Research Funds for the Central Universities under Grant 2022JBXT001 and Grant 2022JBQY004; in part by the National Key Research and Development Program of China under Grant 2020YFB1806903; in part by the National Key Research and Development Program of China under Grant 2021YFB2900301; in part by the National Natural Science Foundation of China under Grant 62221001, Grant 62231009, Grant U21A20445; in part by Fundamental Research Funds for the Central Universities 2022JBQY004; in part by Science and Technology Research and Development Program of China National Railway Group Corporation K2022G018; supported by State Key Laboratory of Advanced Rail Autonomous Operation; supported by Frontiers Science Center for Smart High-speed Railway System; and supported by Beijing Engineering Research Center of High-speed Railway Broadband Mobile Communications. (\emph{Corresponding author: Yong Niu.})
			
			P. Li, H. Wu, G. Sun, Z. Zhong, and B. Ai are with the
			State Key Laboratory of Advanced Rail Autonomous Operation, Beijing Jiaotong University, Beijing 100044, China (e-mail: 19111023@bjtu.edu.cn; hwu@bjtu.edu.cn; guiqisun@bjtu.edu.cn; zhdzhong@bjtu.edu.cn; aibo@ieee.org).
			
			Y. Niu is with the State Key Laboratory of Advanced Rail Autonomous Operation, Beijing Jiaotong University, Beijing 100044, China, and also with the National Mobile Communications Research Laboratory, Southeast University, Nanjing 211189, China (e-mail: niuy11@163.com).
			
			Z. Han is with the Department of Electrical and Computer Engineering at the University of Houston, Houston, TX 77004 USA, and also with the Department of Computer Science and Engineering, Kyung Hee University, Seoul, South Korea, 446-701. (e-mail: hanzhu22@gmail.com).
		
			N. Wang is with the School of Information Engineering, Zhengzhou University, Zhengzhou 450001, China (e-mail: ienwang@zzu.edu.cn).}}

	\maketitle
	
	\begin{abstract}
		One technology that has the potential to improve wireless communications in years to come is integrated sensing and communication (ISAC). In this study, we take advantage of reconfigurable intelligent surface's (RIS) potential advantages to achieve ISAC while using the same frequency and resources. Specifically, by using the reflecting elements, the RIS dynamically modifies the radio waves' strength or phase in order to change the environment for radio transmission and increase the ISAC systems' transmission rate. We investigate a single cell downlink communication situation with RIS assistance. Combining the ISAC base station's (BS) beamforming with RIS's discrete phase shift optimization, while guaranteeing the sensing signal, The aim of optimizing the sum rate is specified. We take advantage of alternating maximization to find practical solutions with dividing the challenge into two minor issues. The first power allocation subproblem is non-convex that CVX solves by converting it to convex. A local search strategy is used to solve the second subproblem of phase shift optimization. According to the results of the simulation, using RIS with adjusted phase shifts can significantly enhance the ISAC system's performance.
	\end{abstract}
	
	\begin{IEEEkeywords}
		Integrated Sensing and Communication (ISAC), mmWave band,  RIS-assisted, High-Speed Railway (HSR).
	\end{IEEEkeywords}
	
	\section{Introduction}
	\IEEEPARstart{I}{N} the upcoming wireless network generation, sensing services will be ubiquitous, such as the Internet of Vehicles, smart home, intelligent manufacturing, human-computer interaction, etc.\cite{ref1}. The requirements for the quality of wireless connections are also increasing. The trend towards bandwidth communication and high-density sensing has resulted in wireless spectrum resources being more restricted. In order to solve this problem, integrated sensing and communication (ISAC) have received more and more attention. ISAC indicates a new type of information processing technology. It can realize the coordination of sensing and communication functions through software and hardware resource sharing or information sharing. By integrating the communication and sensing functions into a single system, ISAC aims to significantly increase energy and spectrum efficiency while lowering hardware and signal expenses. The integration of communications and sensing services, as well as the pursuit of trade-offs and shared performance benefits, are ISAC's ultimate objectives. And ISAC can even pursure greater integration patterns. In this mode, the two functions are designed not just to coexist, rather for the advantage of both parties, i.e., communication-aided sensing or sensing-aided communication.
	
	With the increasing mileage of high-speed rail running, it is particularly important to ensure the safety of the train. Foreign objects intrusion is one of the factors that seriously threaten the railway safety. The use of radar sensing enables the detection of foreign objects around the track and on the train, and some other functions such as sensing the position of the train. However, as the mileage of high-speed railway operations continues to increase, achieving the full railway sensing is a significant challenge in terms of hardware costs. Moreover, in the era of smart railways, the demand for high-capacity connectivity is rising for services like high-definition video monitoring and railway Internet of Things (IoT). Applicating ISAC to the high-speed railway (HSR) system can greatly reduce the hardware costs. It can also solve the problem of tight wireless spectrum resources in the railway private networks.
	
	With its capacity to change wireless communication conditions, reconfigurable intelligent surface (RIS) has received a lot of interest recently. A huge number of passive reflection elements make up the parallel theme known as RIS, every one of which may reflect the received signal with a programmable phase shift. RIS may efficiently enhance the expected signal or suppress the unexpected signal by continuously or discretely modifying the incident signal's amplitude, phase, or both\cite{ref2}.
	
	The application of RIS to ISAC can also serve as reciprocal. One the one hand, the wireless channel connecting the transmitter and the receiver/target is always a determining factor in the sensing and communication performance of ISAC. Sensing targets without or with poor light of sight (LoS) links is a challenging problem. In this case, RIS can be used to provide additional reflective links, improving the sensing accuracy and the transmission rates. On the other hand, a key challenge for wireless communication systems with RIS assistance is the need for perfect channel state information (CSI), which in HSR systems is challenging. If the sensing function of ISAC is used to measure parameters at the receiver (such as the angle of departure/ arrival), it will contribute to the passive beamforming of RIS.
	
	This research examines the communication issue between the ground base station (BS) and the HSR. We installed several mobile relays (MRs) as communication relays on the train's roof since wireless signals can hardly get through the train body. The MRs receive wireless signals from the BS and serve passengers inside the carriages. Trackside BS have both communication and sensing functions, and utilize RIS to raise the system's efficiency. increase network transmission speed with guaranteeing the transmitting power and sensing efficiency thresholds. The optimization variables are composed of the continuous transmitting power of the BS and a finite number of discontinuous RIS phase shifts. They are coupled in the defined optimization objective. The problem is challenging to be solved, and therefore we decompose it into two subproblems and subsequently resolve them..
	
	The following is a summary of the paper's contributions.
	\begin{itemize}
		\item{We construct an HSR-to-ground communication network in which the trackside BS combines the roles of sensing and communication, and we deploy a RIS to offer reflecting pathways to enhance the sensing and communication effectiveness.}
		\item{By maximizing the ISAC BS's transmitting power and the discrete phase shifts of the RIS elements, the system sum rate maximization problem is formulated within the minimal sensing threshold and the maximum transmitting power limitation of the BS. The coupling between the optimization variables makes it challenging to find a closed solution. To do this, we decompose it into two seperate issues and use an iterative alternating optimization approach to solve them.}
		\item{The waveform design subproblem is non-convex, and we convert it into a convex one by employing the first-order Taylor expansion, and then solve it with the CVX toolbox. The RIS phase shift design subproblem is solved by using a local search method to identify the appropriate phase shifts combination.}
		\item {We contrast the suggested algorithm's performance with that of the solution achieved without RIS support and other benchmark methods. In comparison to the other three benchmark schemes, the simulation results demonstrate that the proposed method is able to obtain a higher sum rate under the same sensing signal threshold and other system parameters.}
	\end{itemize}

	The remainder of the paper is structured as follows. In Section~\ref{section: Related Work}, we provide an overview of the relevant works. In Section~\ref{section: System Model}, we create a model that details the system under investigation as well as its communication links. In Section~\ref{section: Problem Formulation}, the system sum rate maximization issue is defined, together with its two subproblems—beamforming design subproblem and phase shift design subproblem. Then Section~\ref{section: Design of RIS-Assisted ISAC System} tailors the optimization techniques to each of the two subproblems.  the optimization algorithms are tailored to each of the two subproblems. Then the two subproblems are jointly optimized to achieve the optimal solution. In Section~\ref{section: Performance Evaluation}, we evaluate the proposed algorithm's performance and contrast it with the various benchmarking algorithms. Finally, Section~\ref{section: Conclusion} serves as the paper's conclusion.
	
	\textbf{\textit{Notations}}: We represent column vectors with boldfaced lowercase letters, e.g., $\mathbf{x}$, and boldfaced uppercase letters represent matrices, e.g., $\mathbf{X}$. We let $\mathbf{X}^T$ and $\mathbf{X}^H$ denote the transpose and Hermitian operations performed on $\mathbf{X}$, respectively. $\mathbb{C}^n$ and $\mathbb{R}^n$ stand for the set of n-dimensional complex and real vectors. $\left [  \mathbf{X}\right ]_i$, $\left [  \mathbf{X}\right ]_{i,j}$ means the $i$-th element of a vector x and the ($i,j$)-th element of a matrix $\mathbf{X}$. $diag(\mathbf{x})$ is a diagonal matrix whose diagonal elements are extracted from a vector x. While the Frobenius norm of a matrix $\mathbf{X}$ is denoted by $\left\| \textbf{X}\right\|_F$. $\bigotimes $ represents the Kronecker product of two matrices.  Finally, we represent a set in flower letters, e.g, $\mathcal{X}$.
	
	\section{Related Work} \label{section: Related Work}
 Tan \textit{et al.} \cite{ref3} discussed novel applications, the major performance requirements, limitations, and future research objectives for ISAC design in 6-th generation mobile networks (6G). Zhang \textit{et al.} \cite{ref4} provided an extensive review of the latest technologies in ISAC systems from a signal processing standpoint, including three types: communication-centric, radar-centric, as well as cooperative design and optimization. Wang \textit{et al.} \cite{ref5} examined the ISAC enabling technologies, such as transmission waveform designing,  signal processing, data processing, environmental modeling, and sensing sources. Xiong \textit{et al.} \cite{ref6} summarized the research status of ISAC waveform design. Hua \textit{et al.} \cite{ref7} studied transmission beamforming in downlink ISAC systems, and in particular, they considered two types of communication receivers, which differed in their ability to eliminate interference from a priori known specialized radar signals. Liu \textit{et al.} \cite{ref8} considered the omnidirectional and directional beampattern design of DFRC downlink communication, and further considered the weighted optimization for the flexible balance between communication and sensing quality based on the solved waveform closed solution. Liu \textit{et al.} \cite{ref9} used hybrid analog-to-digital (HAD) beamforming to create a transceiver design and frame structure for the mmWave dual-function radar communication (DFRC) BS. Islam \textit{et al.} \cite{ref10} provided a paradigm for collaborative optimization while developing analog/digital (A/D) transmission and reception beamformers. Liu \textit{et al.} \cite{ref11} studied the fundamental limitations of ISAC to grasp the mismatch between the restrictions and the existing technology.  In order to improve ISAC's system performance, \cite{ref12, ref13, ref14, ref15 } conducted relevant research. Ouyang \textit{et al.} \cite{ref12} analyzed how well an uplink ISAC system performs and derived new expressions to describe the likelihood of an outage, the traversal communication capacity, and the perceived rate. Li \textit{et al.} \cite{ref13} investigated the  feasibility of spatially-spread orthogonal time frequency space (SS-OTFS) modulation in the ISAC system. Liu \textit{et al.} \cite{ref14} characterized the effectiveness of ISAC by deriving the expressions of precise and asymptotic outage probability (OP), diversity order, approximate ergodic sum rate (ESR). Dong \textit{et al.} \cite{ref15} presented the idea of detecting quality of service (QoS) according to different applications, based on which   a common structure is established for allocating ISAC resources. However, in the above study, the effectiveness limit of ISAC systems was related to the wireless channel’s quality. If the wireless propagation environment was improved, the system performance can be further improved. While none of the above studies took this into account.

RIS is an emerging panel composed of metamaterials that can be used to assist in wireless communications. It consists of a large number of elements much smaller in size than the wavelength, without power amplifiers, and at a very low cost. Based on this, the majority of scholars want to use RIS to enhance the communication quality inexpensively. Renzo \textit{et al.} \cite{ref16} introduced the emerging research area of RIS including wireless communication, proposed a framework for analyzing and optimizing RIS, and summarized the current status of RIS research as well as key research questions. Pan \textit{et al.} \cite{ref17} examined the role of RIS in the future 6G systems and suggested eight possible research directions. Tang \textit{et al.} \cite{ref18} developed a free-space path loss model for RIS-assisted communication for different scenarios, taking into account the distance, RIS panel size, and other factors. Dai \textit{et al.} \cite{ref19} verified the gain of a RIS panel with 256 elements for the low frequency 2.3 GHz and millimeter wave bands, respectively. Nadeem, Tang, Liu \textit{et al.} \cite{ref20,ref21,ref22} studied the sum rate improvement of RIS for MISO, MIMO, and Multiuser MIMO scenarios, respectively. Huang \textit{et al.} \cite{ref23} investigated the energy efficiency of RIS-assisted downlink multi-user communication with the aim of minimizing the link budget of individual users. Guo \textit{et al.} \cite{ref24} maximized all users’ weighted sum rate for multi-user multiple-input single-output (MISO) systems using RIS for both perfect and imperfect channel state information scenarios. Xing \textit{et al.} \cite{ref25} proposed a novel water filling algorithm design method, which can be used to deal with the constraints of various performance indicators and power in imperfect CSI systems. Wei \textit{et al.} \cite{ref26} gave a channel estimation framework for RIS-assisted cascade channels. Basar \textit{et al.} \cite{ref27} took RIS assistance for wireless communication deeper into the modulation phase and proposed the RIS-space shift keying (RIS-SSK) and RIS-spatial modulation (RIS-SM) schemes. Hou \textit{et al.} \cite{ref28} verified the performance gain of RIS for non-orthogonal multiple access (NOMA) systems. All the above studies considered the use of continuous RIS phase shifts. However, from the hardware point of view, the phase shift of the RIS is achieved by the switching of the PIN diodes, so the number of quantized bits of the phase shift is an integer multiple of 2. In fact, a discrete finite phase shift is more reasonable. Di \textit{et al.} \cite{ref29} considered the discrete RIS reflection phase shifts. Zhang \textit{et al.} \cite{ref30} investigated the effect of finite phase shifts on throughput and gave the number of phase shift quantization required for a given throughput constraint. Yang, Chen \textit{et al.} \cite{ref31,ref32} increased the number of RISs in the system to enhance spatial reuse. Yang \textit{et al.} \cite{ref33} used a new active RIS, which not only reflected the signal but also amplified it compared to the passive RIS. There is also a lot of research on the integration of RIS with other technologies. Li \textit{et al.} \cite{ref34} used RIS to solve the UAV LoS link deterioration problem and jointly optimized the UAV trajectory and RIS reflection phase shift matrix to maximize the UAV communication system capacity. Huang \textit{et al.} \cite{ref35} used deep reinforcement learning to jointly optimize transmitter-side beamforming and RIS passive beamforming.

 In order to overcome the dependence of ISAC systems on wireless propagation environments, there have been some studies to apply RIS to ISAC to enhance the communication and sensing performance. Hu \textit{et al.} \cite{ref36} built an ISAC system based on distributed RIS and designed a detailed workflow, including the transmission protocols, position detection, and beamforming optimization. Maximizing the communication and sensing performance can be achieved by jointly optimizing transmit signal waveforms, sensing waveforms, and RIS phase shifts. The main metrics are transmission rate, Cramer-Rao bounded for angle estimation, and sensing mutual information (MI) \cite{ref37, ref38,  ref39, ref40, ref41}. He \textit{et al.} \cite{ref42} and Yu \textit{et al.} \cite{ref43} studied dual-RIS and multiple-RIS assisted ISAC systems, respectively. Zhang \textit{et al.} \cite{ref44} and Sankar \textit{et al.} \cite{ref45} studied the reflection and amplification of active-RIS and hybrid-RIS on communication and sensing signals, respectively. Salem \textit{et al.} \cite{ref46} studied the Multi-user MIMO (MU-MISO) physical layer security ISAC systems when eavesdropped by malicious unmanned aerial vehicles (UAVs). By jointly optimizing the radar receiving beamformer, RIS reflection coefficient, and transmit beamforming, the system's achievable confidentiality could be maximized. Liu \textit{et al.} \cite{ref47} proposed an ISAC system supported by the IRS operating in the terahertz (THz) band. Unlike conventional downlinks, Liu \textit{et al.} \cite{ref48} considered an uplink ISAC system where a single-antenna user used distributed semi-passive RIS to transmit signals to a BS with multiple antennas. Two stages made up the transmission cycle according to the defined framework. The distributed semipassive RIS simultaneously performed position sensing and data transfer at each stage. Position sensing and beamforming design techniques that are straightforward and efficient were also suggested. However, in these above studies, some were not true ISAC, but the communication and sensing functions were deployed on the same base station, the signal was still separate. Moreover, RIS only assisted the communication signal, which did not help the sensing signal. In addition, few studies have taken into account poor quality LoS links, which are very typical in HSR systems.

However, in some extremely harsh communication environments, such as mountainous areas, it is not easy to deploy BS with direct link due to the terrain. Or due to weather, natural disasters, etc., the environment changes causing the original direct link to disappear. We want to achieve communication and sensing in a lower cost and more convenient way. In this paper, we study the impact of RIS in extremely harsh wireless communication environments. We try to use the intelligent reflection of RIS to achieve a certain degree of communication and sensing functions to ensure the safe operation of high-speed railway in the absence of a direct link.

	\begin{figure}[t]
	\centering
	\includegraphics[width=3.2in,height=2.2in]{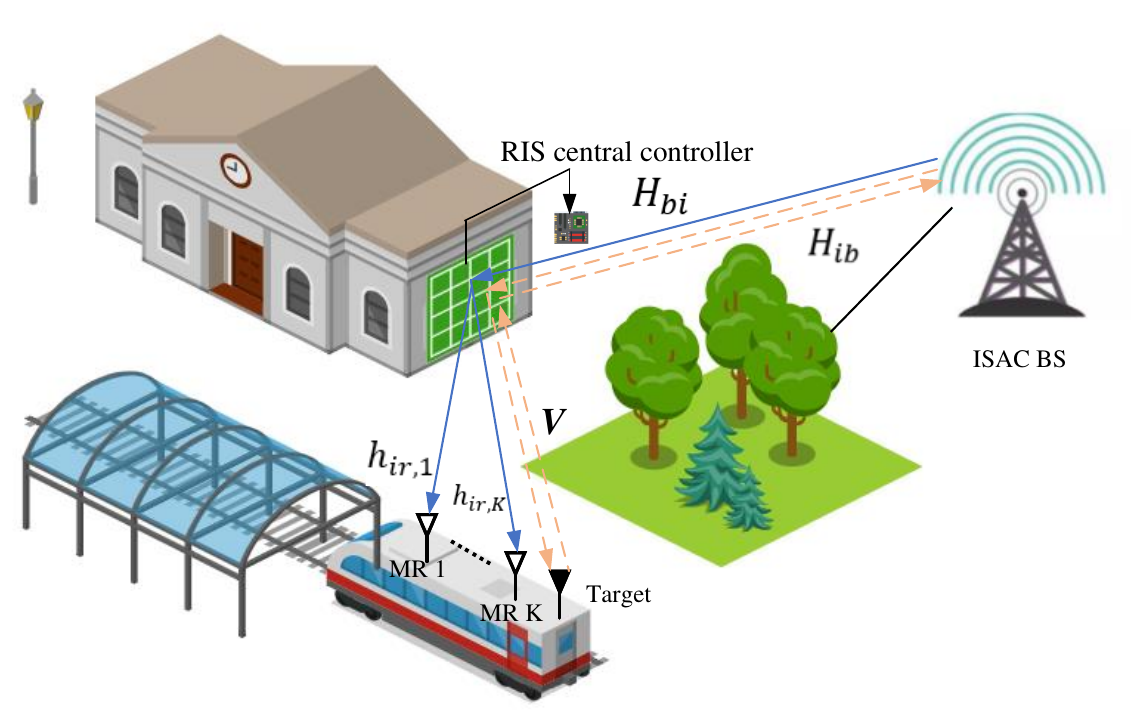}
	\caption{The RIS-aided HSR ISAC system.}
	\label{Fig:System model}
	\label{fig_1}
\end{figure}
	
	\section{System Model} \label{section: System Model}
	\subsection{System Description}
	
As Fig. \ref{Fig:System model} shows that we take into account the downlink of a RIS-assisted ISAC system. An $N$-antenna uniform line array (ULA) ISAC BS is present in the system, which combines sensing and communicating functions. Using the sensing echo signal of the ISAC BS, it can detect illegal targets on the train, tracks, and platforms, and can also achieve high-resolution positioning of carriages or illegal targets. Whether it is detection, estimation, or recognition, the accuracy is determined by the quality of the sensing echo signals. The BS provides services to $K$ single-antenna users, and concurrently detects or tracks targets by receiving echo signals. The direct connection of the ISAC BS to the user and target is disabled due to obstructions such as buildings or vegetation and the weak penetration characteristics of the mmWave. Therefore, we consider deploying a RIS panel in the system to help the signal reach the receivers with the reflection of the RIS. Moreover, the RIS can enhance the reflected signals by adjusting their amplitude and phase through a centralized controller to support communication and radar sensing tasks. The RIS panel with $L$ reflective elements is deployed on the side walls of the building and can maintain good channel conditions between the users and the targets through reasonable deployment. For the sake of simplicity, we make the assumption that the channel is quasi-static flat fading, where the channel fluctuates independently inside the coherent block and stays consistent during the transmission block. Additionally, we presume that the ISAC BS has complete channel information for every connection.

	\subsection{Communication Model}
	
	The signal received by the $k$th MR is:
	
	\begin{equation}
		\label{deqn_ex1a}
		y_k=\boldsymbol h_{ir,k}\boldsymbol{\Phi}\boldsymbol{H}_{bi}\boldsymbol{w}s_k+n,
	\end{equation}

\noindent where $\boldsymbol{H}_{bi}\in \mathbb{C}^{L\times N}$ and $\boldsymbol{H}_{ir}=\left [ h_{ir,1},\cdots h_{ir,K} \right ]^T\in \mathbb{C}^{K\times L}$ represent the coefficients of the ISAC BS-RIS and RIS-MRs channels, respectively, $\boldsymbol{w}\in \mathbb{C}^{N\times 1}$ represents the transmitting beamforming vector, $s_k$ represents the information sent to user $k$, and $n\sim CN(0,\sigma^2)$ is the circularly summetric complex Gaussian (CSCG) noise. $\boldsymbol{\Phi}\triangleq diag\left [ e^{j\phi_1},e^{j\phi_2},\cdots ,e^{j\phi_L} \right ]$ is a diagonal matrix that considers the phase shift that is actually produced about by all RIS components.. $\phi_l=\frac{2m_l\pi}{2^e}$  represents the phase shift of the $l$th element where $l=1,\cdots ,L,m_l\in \left\{ 0,1,\cdots ,2^e-1 \right\}$, and $e$ is the quantity of quantization bits. It is believed that all channels at the ISAC BS are correctly estimated using the inserted pilot symbols. Moreover, all channels are presumed to be perfectly estimated. at the ISAC BS through the inserted pilot symbols. So the signal to interference plus noise ratio (SINR)  that user $k$ has received is:

\begin{equation}
	\label{deqn_ex1a}
	\gamma_k=\frac{\left\|\boldsymbol{h}_{ir,k}\boldsymbol{\Phi H}_{bi} \boldsymbol{w} \right\|}{\sigma^2}.
\end{equation}

All users’ communication sum rate is:

\begin{equation}
	\label{deqn_ex1a}
	\sum_{k\in \mathcal{K}}R_k=\sum_{k\in \mathcal{K}}\frac{C_k}{B}=\sum_{k\in \mathcal{K}}log_2(1+\gamma_k),
\end{equation}

\noindent where $\mathcal{K}$ is the set of all users. The system communication throughput is represented by $C$ and the spectrum's limited bandwidth by $B$, correspondingly.

\subsection{Radar Model}

Considering the terrible situation where The target is obscured from the ISAC BS's line of sight (LoS) connection, we use the RIS to create a virtual LoS link. Using the beampattern gain from the RIS toward the target, we evaluate the efficiency of the sensing system. For a RIS equipped with $L_x$ rows and $L_y$ columns ($L_x\times L_y=L$), its response in the direction of $\theta_a,\theta_e$ is:

\begin{equation}
	\label{deqn_ex1a}
	\boldsymbol{a}\left ( \theta _a,\theta _e \right )\otimes \boldsymbol{a}_z(\theta _e),
\end{equation}

\noindent where,

\begin{equation}
	\label{deqn_ex1a}
	\boldsymbol{a}_y(\theta _a,\theta _e)\triangleq\frac{1}{\sqrt{L_y}}\left [ 1,e^{j\pi sin\theta_acos\theta _e},\cdots ,e^{j\pi (L_y-1) sin\theta_acos\theta _e} \right ]^T,
\end{equation}

\begin{equation}
	\label{deqn_ex1a}
	\boldsymbol{a}_z(\theta _e)\triangleq\frac{1}{\sqrt{L_x}}\left [ 1,e^{j\pi cos\theta _e},\cdots ,e^{j\pi (L_x-1) cos\theta _e} \right ]^T,
\end{equation}

\noindent  and the target's angles contacts with the RIS in the azimuth and elevation planes are denoted by $\theta_a$ and $\theta_e$, accordingly. Then the RIS-to-target beampattern gain is:

	\begin{equation}
	\begin{aligned}
		\mathcal {G}(\theta_a,\theta_e)=&\mathbb{E}\left ( \left|\boldsymbol{a}^H(\theta_a,\theta_e)\boldsymbol{\Phi H}_{bi}\boldsymbol{w} \right|^2 \right )\\
		=&\boldsymbol{a}^H(\theta_a,\theta_e)\boldsymbol{\Phi H}_{bi}\boldsymbol{ww}^H\boldsymbol{H}_{bi}^H\boldsymbol{\Phi}^H\boldsymbol{a}(\theta_a,\theta_e).
	\end{aligned}
\end{equation}

\subsection{Channel Model}

There have been many studies on the estimation of channel state information for RIS-assisted channels, so in this paper it is assumed that the ISAC BS and the RIS central controller have perfect CSI. Suppose there are both LoS and non-light of sight (NLoS) constituents in the system, and all links follow Rician fading. In this work, we suppose that the ISAC BS-RIS link's CSI can be accurately predicted. The channel coefficient between the ISAC BSto the RIS can be denoted as Rice fading matrix $\boldsymbol{H}_{bi}$:

\begin{equation}
	\boldsymbol{H}_{bi}=\sqrt{\frac{K_R}{K_R+1}}\boldsymbol{H}_{bi}^{LoS}+\sqrt{\frac{1}{K_R+1}}\boldsymbol{H}_{bi}^{NLoS},
\end{equation}

\noindent where $K_R=4$ is the Rician factor. $\boldsymbol{H}_{bi}^{LoS}\in \mathbb{C}^{L\times N}$ is the LoS component connecting the RIS and the ISAC BS, which is related to the link distance and remains stable in each time slot. And $\boldsymbol{H}_{bi}^{NLoS}\in \mathbb{C}^{L\times N}$ represents the NLoS Rayleigh fading component. $\boldsymbol{H}_{bi}^{LoS}$ and $\boldsymbol{H}_{bi}^{NLoS}$ can be expressed, respectively, as:

\begin{equation}
	\begin{aligned}
			\left [ \boldsymbol{H}_{bi}^{LoS} \right ]_{l,n}=\sqrt{\beta_0(d_{bi})^{-\alpha_1}}e^{-j\psi_{l,n}},\\
			l\in \left\{1,\cdots,L \right\},n\in \left\{1,\cdots,N \right\},
	\end{aligned}
\end{equation}

\begin{equation}
	\boldsymbol{H}_{bi}^{LoS}=\sqrt{\beta_0(d_{bi}^{-\alpha_2})}\widetilde{\boldsymbol{H}}_{bi}^{NLoS},
\end{equation}

\noindent where $\beta_0=-61.3849$dB indicates the path loss at a distance of $1$m, $d_{bi}$ is the distance between the ISAC BS and RIS, $\alpha_1=2.5$, $\alpha_2=3.6$ are the exponent of path loss in LoS and NLoS scenarios, $\psi_{l,n}$ is the randomly distributed phase within $[0,2\pi)$, and each term of $\widetilde{\boldsymbol{H}}_{bi}^{NLoS}$  is a complicated, circularly-symmetrical, zero-mean, unit-variance random variable to characterize small-scale fading.

Similarly, the channel $\boldsymbol{h}_{ir,k} \in \mathbb{C}^{1\times L}$  of RIS-MR $k$ may be written as:

\begin{equation}
	\boldsymbol{h}_{ir,k}=\sqrt{\frac{K_R}{K_R+1}}\boldsymbol{h}_{ir,k}^{LoS}+\sqrt{\frac{1}{K_R+1}}\boldsymbol{h}_{ir,k}^{NLoS}.
\end{equation}

	\section{Problem Formulation} \label{section: Problem Formulation}
	In this section, using the aforementioned system model, we start by stating the optimization problem, and to effectively find a sub-optimal solution, we divide the challenging optimization problem into two smaller issues.
	
	\subsection{Problem Formulation}
	
	This paper's objective is to simultaneously optimize the ISAC BS beamforming vector along with the RIS phase shift matrix to maximize the system throughput of communication while ensuring the effectiveness of radar sensing. The radar performance can be characterized by the beampattern gain of the RIS. The optimization issue can be described as follows:
	
	\begin{equation}
		\begin{aligned}
			\hspace{0.3cm}&\underset{\boldsymbol w,\boldsymbol \Phi}{\textup{max}}\sum_{k\in \mathcal{K}}\log_{2}\left ( 1+\frac{\left\| \boldsymbol{h}_{ir,k}\boldsymbol{\Phi H}_{bi}\boldsymbol{w}\right\|^2}{\sigma ^2} \right )\\
			s.t.\hspace{0.3cm}&(a) \underset{k\in \mathcal{K}}{\min}\boldsymbol{a}^H(\theta_a,\theta_e)\boldsymbol{\Phi H}_{bi}\boldsymbol{ww}^H\boldsymbol{H}_{bi}^H\boldsymbol{\Phi}^H\boldsymbol{a}(\theta_a,\theta_e)\geq \gamma_{th},\\			
			&(b)\left\|\boldsymbol{w}_2^2 \right\|\leq P_{max},\\
			&(c)\left [ \boldsymbol \Phi \right ]_{l,l}=e^{j\phi_l},\phi_l=\frac{2m_l\pi}{2^e-1},\\
		\end{aligned}
	\end{equation}
	
	\noindent where constraint ($a$) indicates that the minimum beampattern gain of the received signal by the radar detector must be greater than the threshold $\gamma_{th}$; constraint ($b$) is the maximum ISAC BS transmitting power, and $\boldsymbol{w}$ is the beamforming vector; and constraint ($c$) represents the RIS elements' discrete phase shift.
	\subsection{Problem Decomposition}
Obviously, the optimization problem in (12) is non-convex that contains two optimization variables, the beamforming parameter $\boldsymbol{w}$ and the RIS phase shift matrix $\boldsymbol{\Phi}$. During the optimization process of maximizing the communication sum rate, the two optimization variables $\boldsymbol{w}$ and $\boldsymbol{\Phi}$ are coupled to each other, consequently, simultaneous optimization is challenging. Thus, we take into account the alternating optimization procedure. Firstly, optimize $\boldsymbol{w}$ with fixed $\boldsymbol{\Phi}$, then optimize $\boldsymbol{\Phi}$ with fixed $\boldsymbol{w}$, and finally find the optimal $\boldsymbol{w}$ and $\boldsymbol{\Phi}$ that meet the constraints such that the communication sum rate is maximum.
	
	\emph{1) Optimization of $\boldsymbol{w}$  for Given $\boldsymbol{\Phi}$}: The subproblem is to reasonably optimize the ISAC BS transmitting beamforming vector to maximize the communication sum rate based on meeting the minimal SINR constraint of the radar echo signal and the maximum transmitting power constraint. When another optimization variable $\boldsymbol{\Phi}$ is fixed, the optimization issue in (12) can be rewritten as the new form below about the transmitting beamforming vector $\boldsymbol{w}$:
	
	\begin{equation}
	\begin{aligned}
		\hspace{0.3cm}&\underset{\boldsymbol w}{\textup{max}}\sum_{k\in \mathcal{K}}\log_{2}\left ( 1+\frac{\left\| \boldsymbol{h}_{ir,k}\boldsymbol{\Phi H}_{bi}\boldsymbol{w}\right\|^2}{\sigma ^2} \right )\\
		s.t.\hspace{0.3cm}&(a) \underset{k\in \mathcal{K}}{\textup{min}}\boldsymbol{a}^H(\theta_a,\theta_e)\boldsymbol{\Phi H}_{bi}\boldsymbol{ww}^H\boldsymbol{H}_{bi}^H\boldsymbol{\Phi}^H\boldsymbol{a}(\theta_a,\theta_e)\geq \gamma_{th},\\			
		&(b)\left\|\boldsymbol{w}_2^2 \right\|\leq P_{max}.\\
	\end{aligned}
\end{equation}
	
	\emph{2) Optimization of $\boldsymbol{\Phi}$ for Given $\boldsymbol{w}$}: When $\boldsymbol{w}$ is fixed, the optimization problem can be transformed into selecting the best phase shift from $2^e$ phase shifts for maximizing the communication sum rate. Then, the optimization subproblem is formulated as follows:
	
	\begin{equation}
	\begin{aligned}
		\hspace{0.3cm}&\underset{\boldsymbol \Phi}{\textup{max}}\sum_{k\in \mathcal{K}}\log_{2}\left ( 1+\frac{\left\| \boldsymbol{h}_{ir,k}\boldsymbol{\Phi H}_{bi}\boldsymbol{w}\right\|^2}{\sigma ^2} \right )\\
		s.t.\hspace{0.3cm}
		&\left [ \boldsymbol \Phi \right ]_{l,l}=e^{j\phi_l},\phi_l=\frac{2m_l\pi}{2^e-1}.\\
	\end{aligned}
\end{equation}
	
For the above two optimization subproblems, in the next sections, we will construct appropriate optimization strategies to address them, so as to accomplish the objective of boosting communication total rate while fulfilling restrictions.

	\section{Design of RIS-Assisted Wireless ISAC System} \label{section: Design of RIS-Assisted ISAC System}
	
	In this section, we first address the two aforementioned subproblems, followed by a sum rate maximization solution. It alternately handles these two subproblems iteratively until the algorithm converges and a suboptimal result is obtained.
	
	\subsection{Beamforming Design Sub-Problem Algorithm }
	
	To simplify the problem, we denote $\boldsymbol{G}_k=diag(\boldsymbol{h}_{ir,k} )\boldsymbol{H}_{bi}, \boldsymbol{W}=\boldsymbol{ww}^H,\boldsymbol{u}=\left [e^{j\phi_1},\cdots,e^{j\phi_L} \right ]$, and then we obtain the following equation:
	
	\begin{equation}
		\begin{aligned}
			&\underset{\boldsymbol w}{\max}\sum_{k\in \mathcal{K}}\log_{2}\left ( 1+\frac{\left\| \boldsymbol{h}_{ir,k}\boldsymbol{\Phi H}_{bi}\boldsymbol{w}\right\|^2}{\sigma ^2} \right )\\
			=&\underset{\boldsymbol W}{\max}\sum_{k\in \mathcal{K}}\log_{2}\left ( 1+\frac{\textbf{Tr}(\boldsymbol{WG}_k^H\boldsymbol{uu}^H\boldsymbol{G}_k)}{\sigma^2} \right )\\
			=&\underset{\boldsymbol{W}}{\min}\sum_{k\in \mathcal{K}}\left ( F_1(\boldsymbol{W})-\log_{2}(\sigma^2) \right ),
		\end{aligned}
	\end{equation}

\noindent where where $F_1(\boldsymbol{W})\triangleq \log_{2}\left ( \textbf{Tr}\left ( \boldsymbol{WG}_k^H\boldsymbol{uu}^H\boldsymbol{G}_k \right ) +\sigma^2\right )$. Now, we rewrite the initial optimization issue in (13) into an equivalent form to make using the alternating optimization algorithm easier:

	\begin{equation}
	\begin{aligned}
		\hspace{0.3cm}&\underset{\boldsymbol W}{\min}\left ( F_1(\boldsymbol{W})-\log_{2}(\sigma^2) \right )\\
		s.t.\hspace{0.3cm}&(a) \underset{k\in \mathcal{K}}{\textup{min}}\boldsymbol{a}^H(\theta_a,\theta_e)\boldsymbol{\Phi H}_{bi}\boldsymbol{W}\boldsymbol{H}_{bi}^H\boldsymbol{\Phi}^H\boldsymbol{a}(\theta_a,\theta_e)\geq \gamma_{th},\\			
		&(b)\textup{Tr}(\boldsymbol{W})\leq P_{max},\\
		&(c)\boldsymbol{W}\succeq0,\\
		&(d)Rank(\boldsymbol{W})\leq1.
	\end{aligned}
\end{equation}

The problem, known as the difference of concave functions (DC) problem, is still challenging to resolve. To do this, we make an attempt to transform it into a convex function utilizing the Taylor expansion of the first order. That is, in order to estimate the ideal solution from the upper limit, one must first establish a rough upper limit for the optimization problem and minimize it as much as possible. By this approximation, the original goal function can be converted into a convex function with respect to $\boldsymbol{W}$. Specifically, the objective function $F_1(\boldsymbol{W})$ at the $i$-th iteration of the first-order Taylor expansion may be represented as:

\begin{equation}
	\begin{aligned}
	F_1(\boldsymbol{W})=&F_1(\boldsymbol{W}^i)+\textup{Tr}\left ( \triangledown_{ \boldsymbol{W}}F_1(\boldsymbol{W}^i)^H(\boldsymbol{W}-\boldsymbol{W}^i) \right )\\
	\triangleq&\widetilde{F_1}(\boldsymbol{W},\boldsymbol{W}^i),
	\end{aligned}
\end{equation}

\begin{equation}
	\triangledown_{\boldsymbol{W}}F_1(\boldsymbol{W}^i)=\frac{\boldsymbol{G}_k^H\boldsymbol{uu}^H\boldsymbol{G}_k}{\sigma^2ln2}.
\end{equation}
	
Currently, the nonconvexity of the optimization problem in (19) is only from a rank-one constraint in (16$d$). First, we release the non-convex restriction in (16$d$) in order to resolve this problem with SDR and obtain its relaxed version, and then the problem can be solved with the CVX solver directly.

\begin{algorithm}[!t]
	\caption{Successive Convex Approximation-Based Algorithm}\label{alg:alg1}
	\KwIn{the number of quantization bits $e$, the sensing threshold $\gamma_{th}$, convergence threshold $\delta$}
	Initialize  the initial iteration index $i=1$, $\boldsymbol{\Phi}^i$, $\boldsymbol{W}^i$\\
	\For{$R^{i+1}-R^i>\delta$}{
		Solve (16) for specified $\boldsymbol{W}^i$ and record the interim answer $\boldsymbol{W}$\\
		Set $i=i+1$ and $\boldsymbol{W}^i=\boldsymbol{W}$
	}
	\KwOut{{$\boldsymbol{W}^*=\boldsymbol{W}^i$}}
\end{algorithm}

\subsection{Phase Shift Design Sub-problem Algorithm}
	
The initial optimization issue takes on a new form with the fixed transmission beamforming vector, such (14), where only RIS discrete phase shift matrix remains. It is still non-convex, making it challenging to resolve via convex optimization. Considering the complexity of the problem, the simplest straightforward method for solving this non-convex problem is the exhaustive algorithm. However, it is clear that the exhaustive algorithm's complexity rises exponentially as the set of practical solutions increases \cite{ref49}. The temporal complexity of the local search method and the exhaustive search algorithm are contrasted. Fortunately, the local search algorithm has much lower time complexity than the exhaustive algorithm. Therefore, we choose the local search algorithm in Algorithm \ref{alg:alg2}  to solve this optimization subproblem. Especially, we firstly fix the phase shift of the other $L-1$ elements unchanged, then traverse all possible values for each RIS unit $\theta_l$ and select the most suitable value $\theta_l^*$ that satisfies the central objective, that is, optimizing the communication sum rate. The other elements are then optimized in the same manner, until all phase shifts in the array have been entirely optimized.

\subsection{Sum Rate Maximization}

We provide a summary of the discrete phase shift optimization and beamforming vector designing subproblem algorithms discussed above and present a sum rate maximization solution. According to Algorithm \ref{alg:alg3}, we randomly set the initial state and then alternately modify the transmitting beamforming parameters and phase shift matrix as long as the algorithm is not yet convergent, i.e. the system transmission rate differential between two subsequent rounds is less than a specified limit $\left|R^{(i+1)}-R^{(i)} \right|< \min\{\vartheta ,\theta\}$.

\begin{algorithm}[!t]
	\caption{Phase Shift Design Sub-problem Algorithm Local Search-Based Algorithm}\label{alg:alg2}
	\KwIn{the number of quantization bits $e$, the sensing threshold $\gamma_{th}$, convergence threshold $\vartheta $}
	Initialize  the initial iteration index $i=1$, $\boldsymbol{\Phi}^i$, $\boldsymbol{W}^i$\\
	\For{$R^{i+1}-R^i>\vartheta$}{
		Give every possible value to $\phi_l$, and choose the value that maximizes the sum rate in problem (14), denoted as $\phi_l^*$;\\
		$\phi_l=\phi_l^*$;
	}
		\KwOut{{$\boldsymbol{\Phi}^*=\boldsymbol{\Phi}^i$}}
\end{algorithm}

\begin{algorithm}[!t]
	\caption{Sum Rate Maximization}	\label{alg:alg3}
	\KwIn{Thresholds $\gamma_{th}$, $\theta$, $\vartheta $, randomly generate \bm{$\Phi$}, \bm{$\Phi^*$}=\bm{$\Phi$}, phase shift quantization bits $e$, the initial iteration index $i=1$}
	Update $\bm{{\rm W}}$$_{i+1}$ by Algorithm 1 with given \bm{$\Phi$}$_i$ ;\\
	Update \bm{$\Phi$}$_{i+1}$ by Algorithm 2with given $\bm{{\rm W}}$$_i$;\\
	\eIf{$\left | R^{(i+1)}-R^{(i)}\right |<\min\{\vartheta ,\theta\}$}{
		$R^*=R^{(i+1)}$;\\
		\bm{$\Phi^*$}=\bm{$\Phi$}$_{i+1}$;\\
		\bm{${\rm W^*}$}=\bm{${\rm W}}_{i+1}$;\\
		\KwOut{\bm{${\rm W^*}$}, \bm{$\Phi^*$}, $R^*$;}}{
		$i=i+1$, and go to Step 1;
	}
\end{algorithm}
	
	\subsection{Complexity Analysis}
	
The complexity of the method for maximizing the sum rate is not only determined by the number of iterations, which  is configured to $N_{outer}$ for the convergence condition to be met $\left|R^{(i+1)}-R^{(i)} \right|<\min\{\vartheta ,\theta\}$, but also in terms of the two subproblems. For the first, how often gradients are updated and the optimization of beamforming of all MRs in each gradient iteration create complexity. The number of receivers is $M$, and we write $N_{inner}$ for the number of gradient descent iterations. Thereby, the complexity of beamforming design subproblem is $\mathcal{O}(N_{inner}\times M)$. For the latter, the local search algorithm changes the value of element $l$ while keeping the other phase shifts constant and choosing the best one out of $2^e$ possible ones, and updates a value for $\phi_l$. Because the RIS contains $L$ elements, this part's complexity is $\mathcal{O}(L\times2^e)$. Consequently, we obtain $\mathcal{O}(N_{outer}\times(N_{inner}\times M+L\times 2^e))$ for the suggested sum rate maximization algorithm's complexity.

	\section{Performance Evaluation} \label{section: Performance Evaluation}
	In this section, we analyze how well the ISAC system is working with RIS. We get numerical data for various key parameters by comparing with other benchmark methods, which allows us to assess the efficiency of our suggested strategy, as well as analyze the influence of numerous essential aspects on effectiveness.
	
	\subsection{Simulation Setup}
	We take into account the mmWave HSR system. A high-speed train has eight carriages, each measuring 200 meters in length. The roof of the high-speed rail is evenly deployed with 11 MRs, of which each carriage is equipped with 1-2 MRs. A reflective link is provided by deploying a RIS between the ISAC BS and the MRs to assist in the transmission of communication and sensing signals. We construct a three-dimensional Cartesian coordinate system, as seen in Fig. 2. The high-speed rail and the MRs are parallel to the Y axis, while the RIS is set up in the X-Z plane. The ISAC BS and the MRs are located on different sides of the Y-axis. One Target is randomly distributed around 11 MRs. Table \ref{tab:table1} shows the system simulation parameter settings.

		\begin{figure}[!t]
		\centering
		\includegraphics[width=3.2in,height=2.5in]{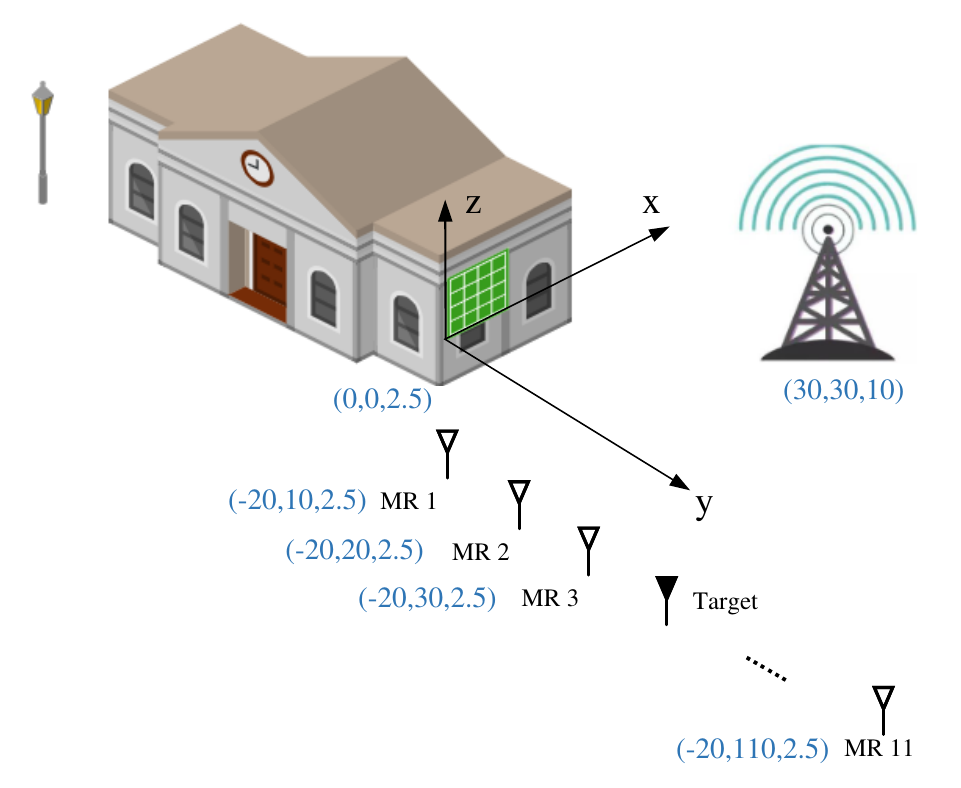}
		\caption{The ISAC system's cartesian coordinate with RIS assistance.}
		\label{Fig:2}
	\end{figure}

	\begin{table}[!t]
		\caption{Simulation Parameters\label{tab:table1}}
		\label{table1}
		\centering
		\begin{tabular}{lll}
			\hline
			Parameter & Symbol & Value\\
			\hline
			Number of ISAC transmit antennas & $N$ & 8\\
			Number of MRs & $K$ & 11\\
			Amount of RIS components & $L$ & 64\\
			Number of phase quantization bits & $e$ & 3\\
			Carrier frequency & $f$ & 30GHz\\
			Bandwidth & $W$ & 100MHz\\
			Maximum transmitting power  & $P$ & 23dBm\\
			Radar SINR threshold & $\gamma_{th}$ & $0.5\times10^{-4}$\\
			Noise power spectral density & $N_0/W$ & -134dBm/MHz\\
			Noise power spectral density & $\sigma^2$ & -134dBm/MHz\\
			Normalized antenna spacing &$\delta$ & $\lambda/2$\\
			Rician factor &$K_R$ & 4\\
			Path loss exponent & $\beta$ & 2\\
			Height of the ISAC BS& $h_{BS}$ & 10m\\
			Height of MRs & $h_{MR}$ & 2.5m\\
			Height of RIS & $h_{RIS}$ & 2.5m\\
			\hline		
		\end{tabular}
	\end{table}
	
	\subsubsection*{\textup{We contrast the proposed approach with the following algorithms to demonstrate how well it performs in the system}}
	\begin{itemize}
		\item{\textbf{Without-RIS:} the technique does not deliver reflected signals via RIS, and the receiver may only receive signals through direct lines. In order to achieve our objectives, utilize the same power allocation algorithm.}
		\item{\textbf{Random Phase Shift (RPS):} for each element of RIS, the scheme chooses a set of functional phase shifts at random and maintains other elements throughout time. The most effective beamforming vector is then determined using algorithm \ref{alg:alg1} with the transmitting power of all links, and the goal of maximizing communication sum rate is explored.}
		\item{\textbf{Average Power Transmission (APT):} in the algorithm, every transmit antenna is given the same amount of transmission power, after which finds the RIS phase shift which optimizes the overall system communication throughput using a local search technique, and discusses the maximum communication sum rate and perception performance.}
	\end{itemize}

	\subsection{Performance Evaluation}

\begin{figure}[!t]
	\centering
	\includegraphics[width=3.2in,height=2.5in]{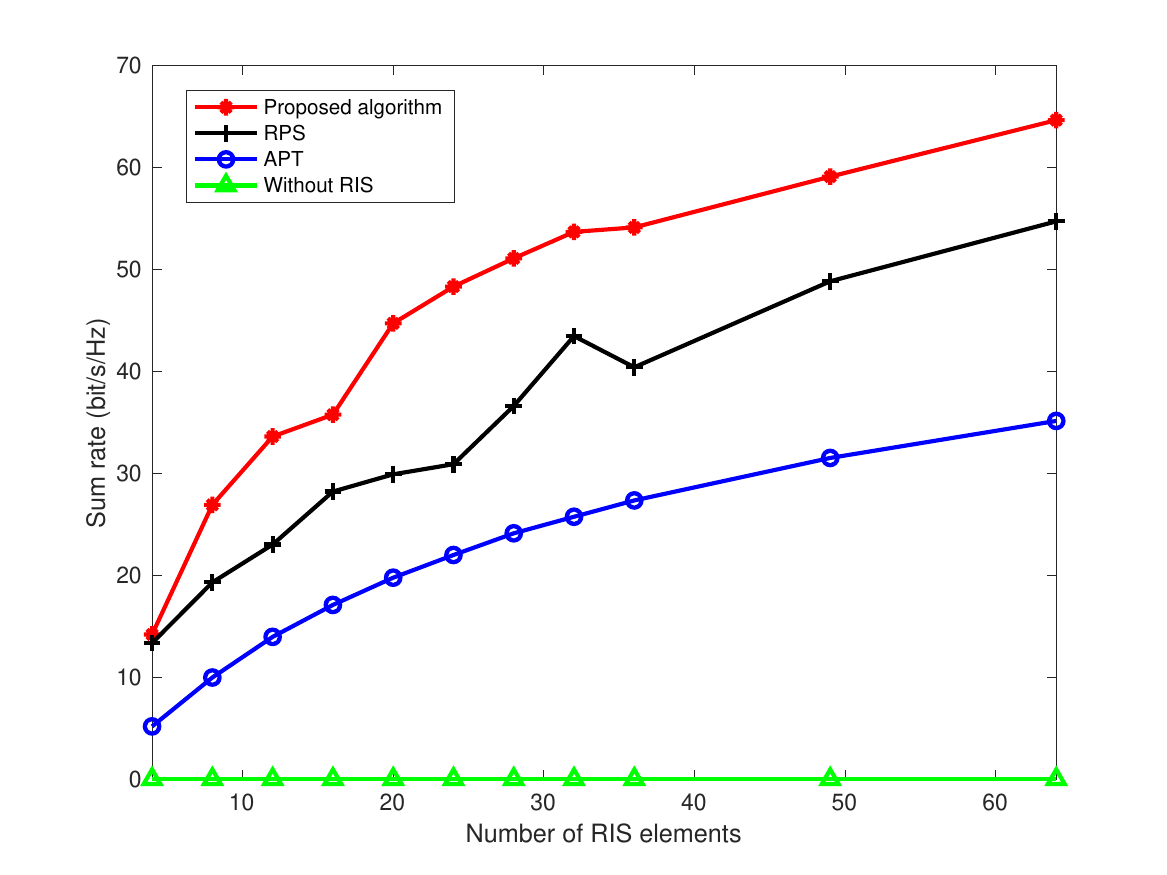}
	\caption{Sum rate versus the quantity of RIS elements $L$.}
	\label{Fig:3}
	\label{fig_1}
\end{figure}

	In Fig. \ref{Fig:3}, we defined the quantity of quantization bits as $e=3$, and the threshold of radar sensing beampattern gain $\gamma_{th}=0.5\times10^{-4}$. We plot the communication sum rate curve as the RIS steadily expanded from 4 to 64 elements. The findings demonstrated that as the quantity of RIS components increases, we can see that the communication sum rate of the three RIS auxiliary schemes all increases. This indicates that more RIS elements can provide more reflective links and thus transmit more data. However, this upward trend gradually slows down with the increase of $L$, which shows that the continuous expansion of the RIS panels cannot increase the sum rate indefinitely with other parameters unchanged. The communication  sum rate provided by the reflective link is limited, which is mainly constrained by the transmission power, Shannon sum rate, system model, physical environment and other factors. The suggested algorithm with RIS has a highest communication sum rate than the previous three comparison algorithms. It shows the RIS’s ability to deliver reflected link auxiliary transmission when the ISAC BS’s direct connection to the MRs is interrupted by the harsh environment and cannot provide service. Compared to the RPS and APT schemes, the proposed scheme performs better, because the random RIS phase shifts do not give full play to the maximum advantage of the RIS elements. The channel variations between various MRs are also not taken into account by the average transmission power. As a result, when the sensing gain threshold is met, the proposed RIS-aided system outperforms the all baseline schemes in terms of communication performance.

\begin{figure}[!t]
			\centering
			\includegraphics[width=3.2in,height=2.5in]{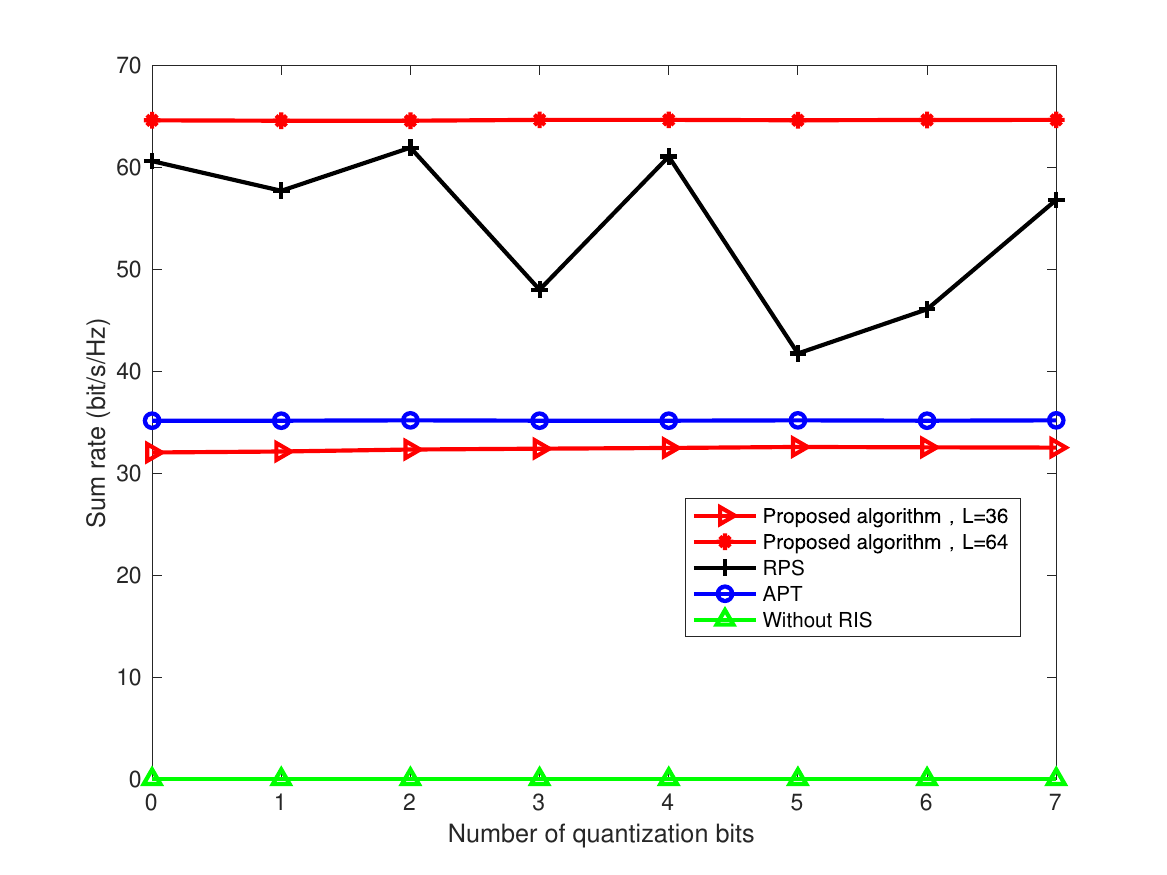}
			\caption{Sum rate versus the quantity of quantization bits $e$.}
			\label{Fig:4}
			\label{fig_1}
\end{figure}
	
\begin{figure}[t]
	\centering
	\includegraphics[width=3.2in,height=2.5in]{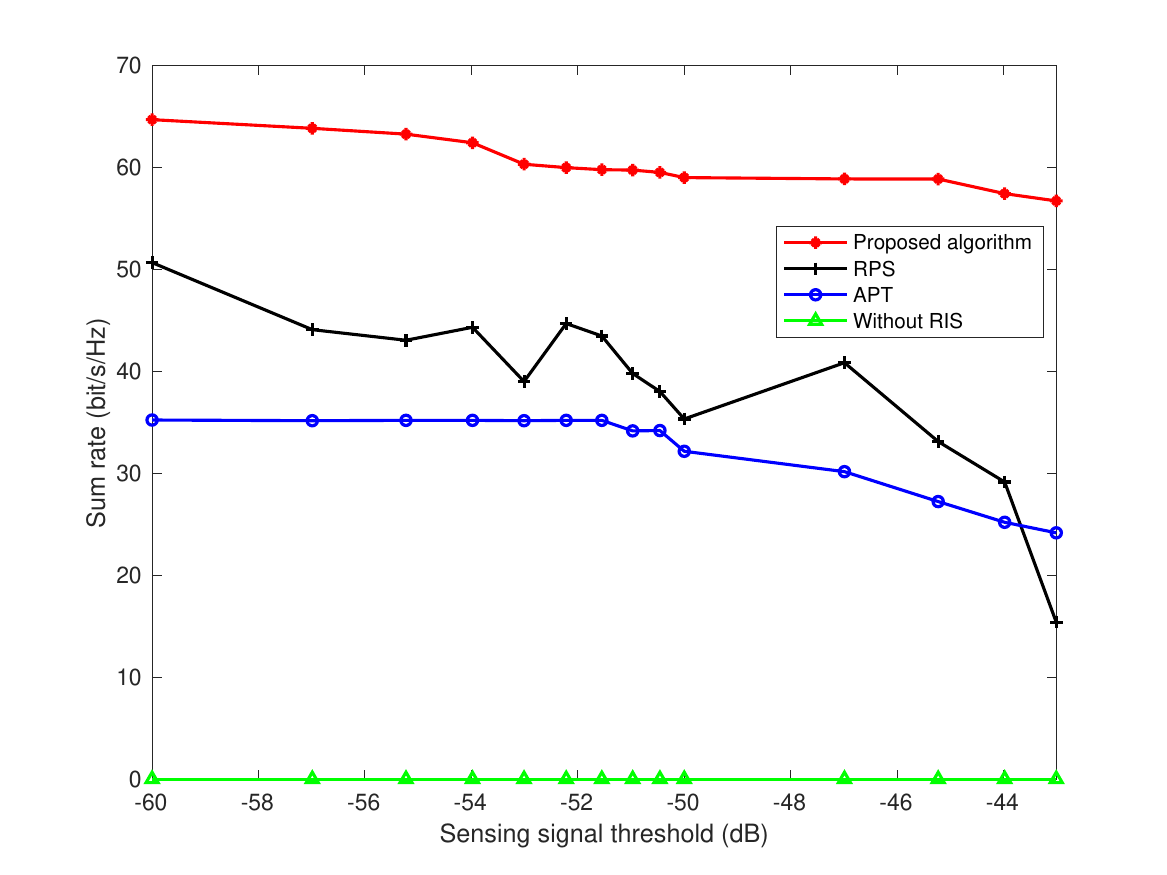}
	\caption{Sum rate versus Sensing signal threshold.}
	\label{Fig:5}
	\label{fig_1}
\end{figure}

\begin{figure}[t]
	\centering
	\includegraphics[width=3.2in,height=2.5in]{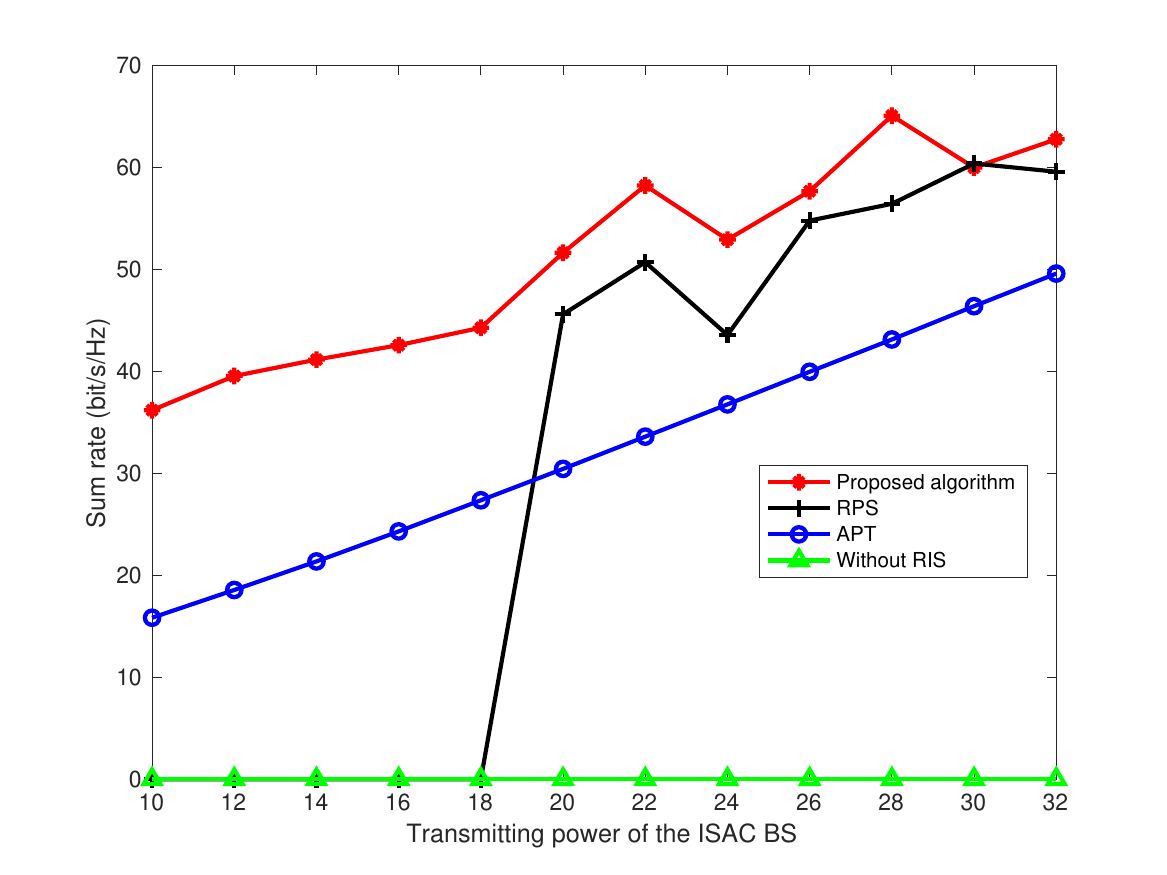}
	\caption{Sum rate versus transmitting power of the ISAC BS.}
	\label{Fig:6}
	\label{fig_1}

\end{figure}
			\begin{figure*}[t]
		\centering
		\subfloat[]{\includegraphics[width=3.2in,height=2.5in]{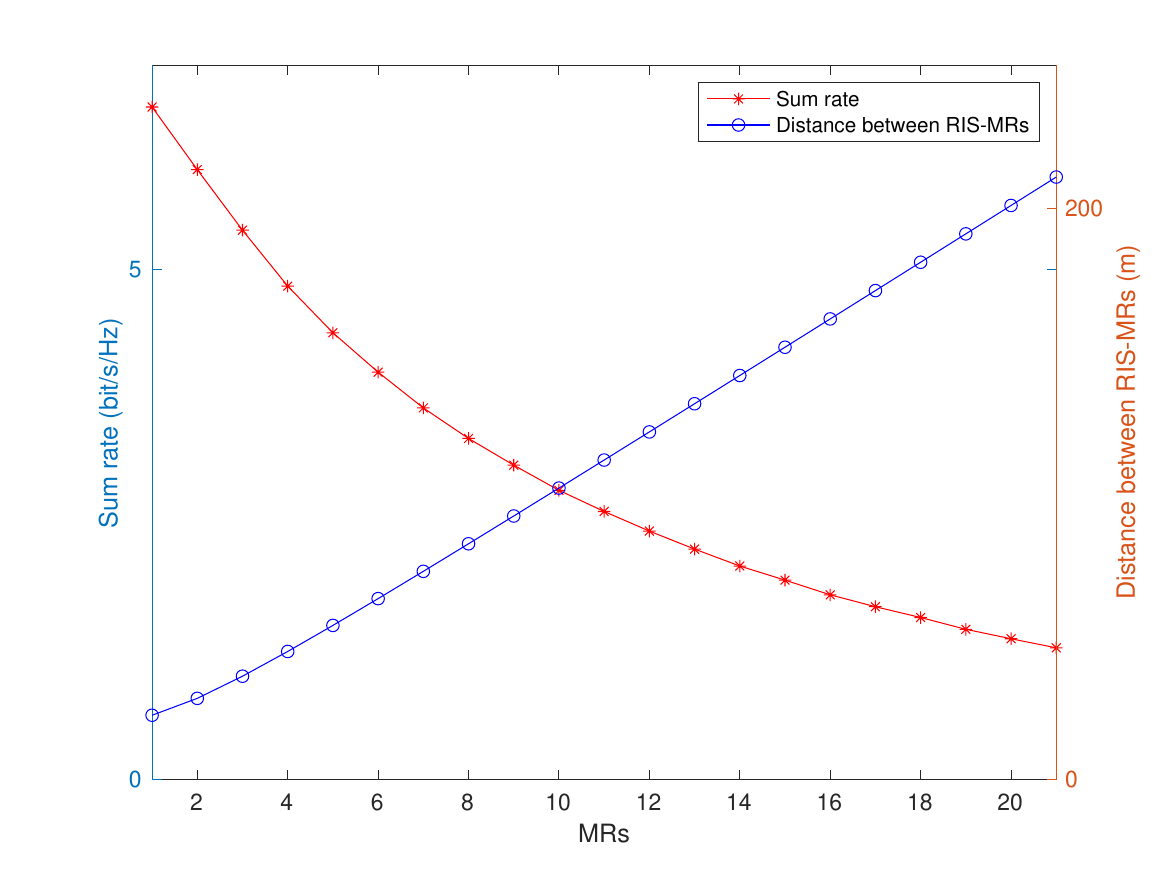}%
			\label{fig_first_case}}
		\hfil
		\subfloat[]{\includegraphics[width=3.2in,height=2.5in]{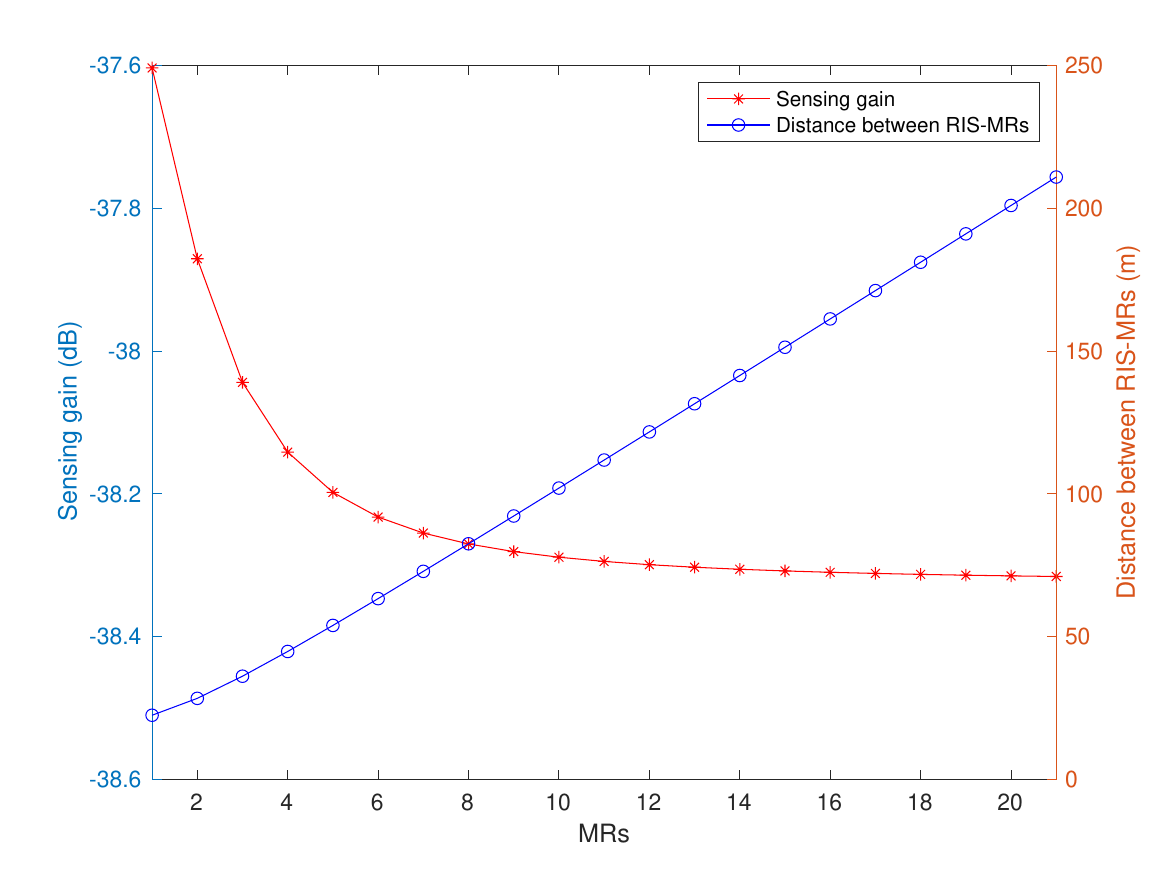}%
			\label{fig_second_case}}
		\caption{(a) Sum rate versus distance between RIS and MRs. (b) Sum rate versus distance between RIS and MRs.}
		\label{Fig:7}
		\label{fig_sim}
	\end{figure*}
	In Fig. \ref{Fig:4}, we set $L=36/64$ and the radar sensing beampattern gain threshold is $\gamma_{th}=0.5\times10^{-4}$. We show the variation in communication efficiency against the quantity of quantized RIS element phase shift bits. As shouwn in the figure, both the proposed scheme and the communication sum rate of APT hardly change with the increase of $e$, while the float of the RPS scheme is large and irregular. This is because regardless of the number of quantization bits, the proposed algorithm and APT algorithm both find the optimal phase shift of each RIS element through the local optimization algorithm, and also consider the cooperation between all RIS elements. Through multiple iterations, the superior performance of the selected RIS phase shift matrix is guaranteed. This also may be due to the fact that the optimization of the transmission beamforming vector has a greater impact on the sum rate than the phase shift of the RIS. Therefore, after power optimization of the proposed scheme and the APT scheme, no matter what the quantization bits of RIS phase shift are, the optimal phase shift matrix can be found to maximize the sum rate. On the other hand, the RPS algorithm does not optimize the RIS phase shifts, but only randomly selects a phase shift for each RIS element, so there is no guarantee that the selected phase shift is suitable. Therefore, the strength of each reflected signal cannot be guaranteed, and the fluctuation is large. Correspondingly, the maximum sum rate cannot be guaranteed. Therefore, the sum rate of RPS fluctuates greatly. This also may be due to the fact that the optimization of the transmission beamforming vector has a greater impact on the sum rate than the phase shift of the RIS. Therefore, after power optimization of the proposed scheme and the APT scheme, no matter what the quantization bits of RIS phase shift are, the optimal phase shift matrix can be found to maximize the sum rate. On the other hand, the RPS algorithm does not optimize the RIS phase shifts, but only randomly selects a phase shift for each RIS element, so there is no guarantee that the selected phase shift is suitable. Therefore, the strength of each reflected signal cannot be guaranteed, and the fluctuation is large. Correspondingly, the maximum sum rate cannot be guaranteed. Therefore, the sum rate of RPS fluctuates greatly.

	In Fig. \ref{Fig:5}, we set $L=64$, and $e=3$. We show the variation in sum rate as a function of the sensing beampattern gain threshold. As illustrated in Fig. \ref{Fig:5}, no matter the proposed algorithm, RPS, or APT, the sum rate rapidly declines as the sensing threshold rises. This is because as the sensing performance requirements for the target detection increase, the transmission power and the RIS phase shifts need to be adjusted to meet the higher threshold of sensing gain, and the communication sum rate will decrease accordingly. Morever, as the sensing gain threshold increases, the decrease in sum rate becomes more obvious. Among of the all three RIS-assisted algorithms, the RPS scheme has the most severe reduction in sum rate. This is because RPS selects RIS phase shifts randomly, and if an inappropriate RIS phase shift matrix is selected, the sensing performance needs to be compensated for by power allocation, and the sum rate will decrease accordingly.

	In Fig. \ref{Fig:6}, we set $L=64$, $e=3$, and the radar sensing beampattern gain is $\gamma_{th}=0.5\times10^{-4}$. We plot the sum rate according to the ISAC BS's highest transmission power. As shown in Fig. \ref{Fig:6}, the sum rate of all RIS-assisted schemes increases with the maximum transmission power. Our suggested method has a substantially higher sum rate than the other three benchmark schemes. It should be noted that when the transmission power $P_{max}<=18dB$, the sum rate of the RPS scheme is 0. This is because when the ISAC BS's transmitting power is small, the randomly selected RIS phase shift matrix cannot meet the sensing gain threshold requirement, then the optimization problem we establish is not feasible, so the sum rate is 0.
	
	In Fig. \ref{Fig:7}, we set $L=64$, $e=3$, and the radar sensing beampattern gain is $\gamma_{th}=0.5\times10^{-4}$. We compare the sum rate and the sensing beampattern gain of all MRs. As shown in Fig. \ref{Fig:7}, from the first MR to the 11st MR, RIS and MRs are becoming further apart, and the path loss is getting worse, so the sum rate and the sensing beampattern gain gradually decrease. Among them, the decrease of the sensing beampattern gain is faster than that of the sum rate, because the sensing signal goes through the ISAC BS-RIS-Target and the Target-RIS-ISAC BS backhaul links. The path loss introduced by the doubling distance is much greater than that of the communication signal. Therefore, in order to ensure that the sensing performance of all possible targets meets the threshold requirements, the setting of the optimization parameters needs to pay more attention to the MR at the farthest location. This also ensures fairness in the sum rate to all MRs. While, the possible sum rate is not maximum correspondingly. This is the cost of balancing communication and sensing performance.

	\section{Conclusion} \label{section: Conclusion}
	In this paper, we have focused on a downlink ISAC system with the aid of a RIS. To improve the system sum rate with meeting the realistic discrete phase shift, power, and sensing QoS limitations, We have looked into a hybrid beamforming architecture and RIS phase shift optimization problem. The specified issue is non-convex, hence we have provided a alternative optimization procedure to arrive at a sub-optimal solution. Finally, the proposed algorithm's superiority has been confirmed by numerical results, and we have seen that RIS may improve the condition of a channel by giving numerous controlled signal reflections depending on how many RIS elements there are. In the future work, we will consider RIS-assisted multi-cell HSR integrated sensing and communication to improve the handover success rate and enhance the practicability.

	\begin{IEEEbiography}[{\includegraphics[width=1in,height=1.25in,clip,keepaspectratio]{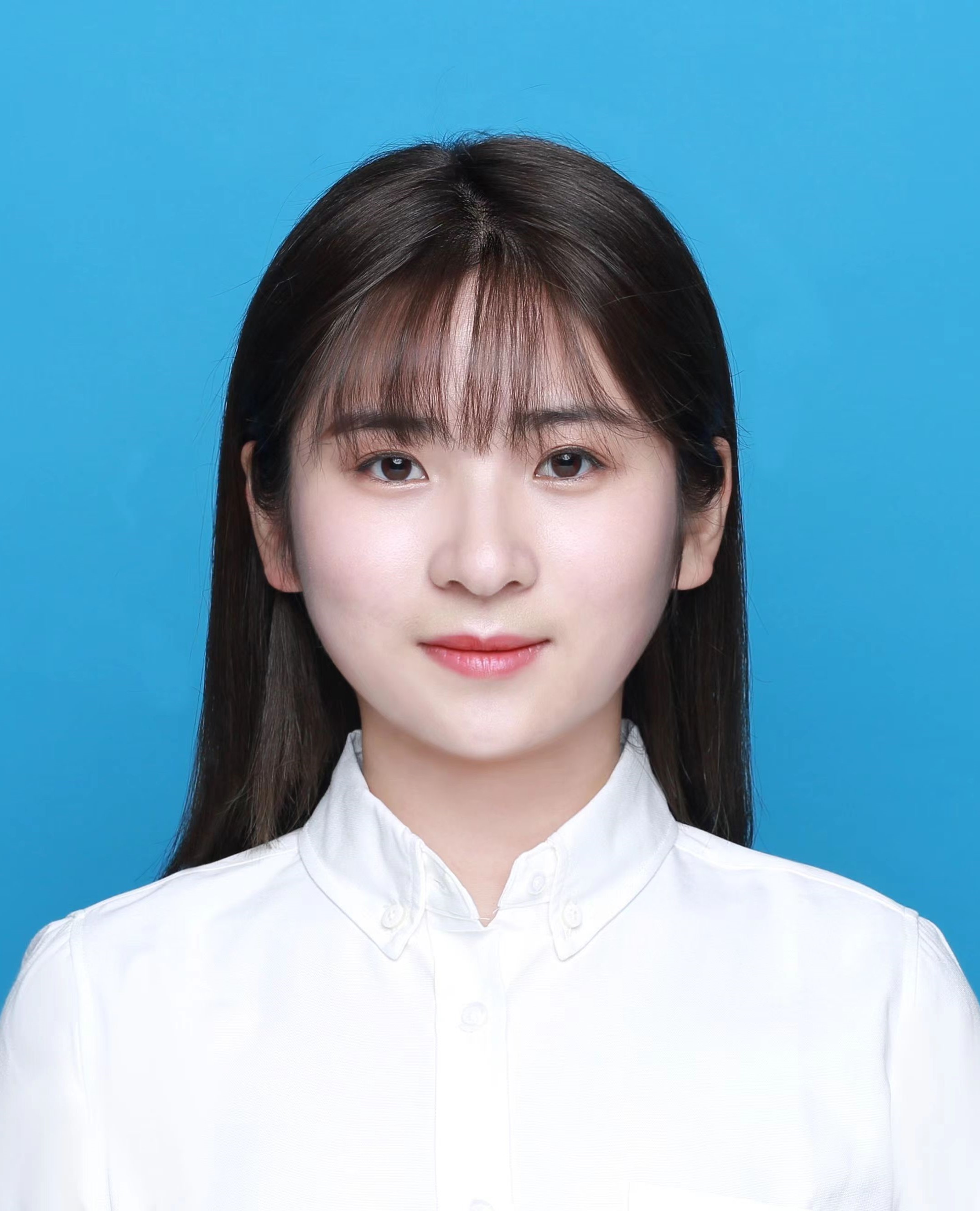}}]{Panpan Li}
		 was born in Henan, China, in 1997. She received the B.S. degree in communication engi- neering from North China Electric Power University, Baoding, China, in 2019. She is currently pursuing the Ph.D. degree with the State Key Laboratory of Rail Traffic Control and Safety, Beijing Jiao- tong University, Beijing, China. Her current research includes millimeter-wave wireless communications, reconfigurable intelligent surfaces, convex optimiza- tion and wireless resource allocation.
		
	\end{IEEEbiography}
	\vspace{-10mm}

	\begin{IEEEbiography}[{\includegraphics[width=1.4in,height=1.25in,clip,keepaspectratio]{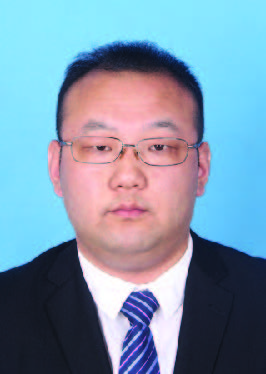}}]{Yong Niu}
		(M’17) received the B.E. degree in Electrical Engineering from Beijing Jiaotong University, China, in 2011, and the Ph.D. degree in Electronic Engineering from Tsinghua University, Beijing, China, in 2016.
		
		From 2014 to 2015, he was a Visiting Scholar with	the University of Florida, Gainesville, FL, USA. He is currently an Associate Professor with the State Key Laboratory of Advanced Rail Autonomous Operation, Beijing Jiaotong University. His research interests are in the areas of networking and communications, including millimeter wave communications, device-to-device communication, medium access control, and software-defined networks. He received the Ph.D. National Scholarship of China in 2015, the Outstanding Ph.D. Graduates and Outstanding Doctoral Thesis of Tsinghua University in 2016, the Outstanding Ph.D. Graduates of Beijing in 2016, and the Outstanding Doctorate Disserta- tion Award from the Chinese Institute of Electronics in 2017. He has served as Technical Program Committee member for IWCMC 2017, VTC2018-Spring, IWCMC 2018, INFOCOM 2018, and ICC 2018. He was the Session Chair for IWCMC 2017. He was the recipient of the 2018 International Union of Radio Science Young Scientist Award.
	\end{IEEEbiography}
	\vspace{-20mm}

	\begin{IEEEbiography}[{\includegraphics[width=1in,height=1.25in,clip,keepaspectratio]{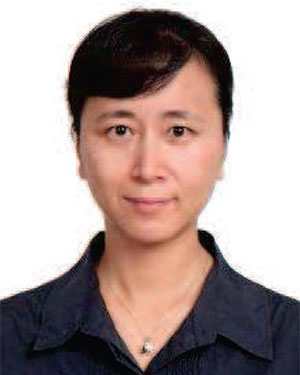}}]{Hao Wu}
		received the Ph.D. degree in information and communication engineering from the Harbin Institute
		of Technology, Harbin, China, in 2000. She is currently a Full Professor with the State Key Laboratory of Advanced Rail Autonomous Operation, Beijing Jiaotong University, Beijing, China. She has authored or coauthored more than 100 papers in international journals and conferences.Her research interests include intelligent transportation systems, security and QoS issues in wireless networks (VANETs, MANETs, and WSNs), wireless communications, and Internet of Things. She is a Reviewer of major conferences and journals in wireless networks and
		security for IEEE.
	\end{IEEEbiography}

	\begin{IEEEbiography}[{\includegraphics[width=1in,height=1.15in,clip,keepaspectratio]{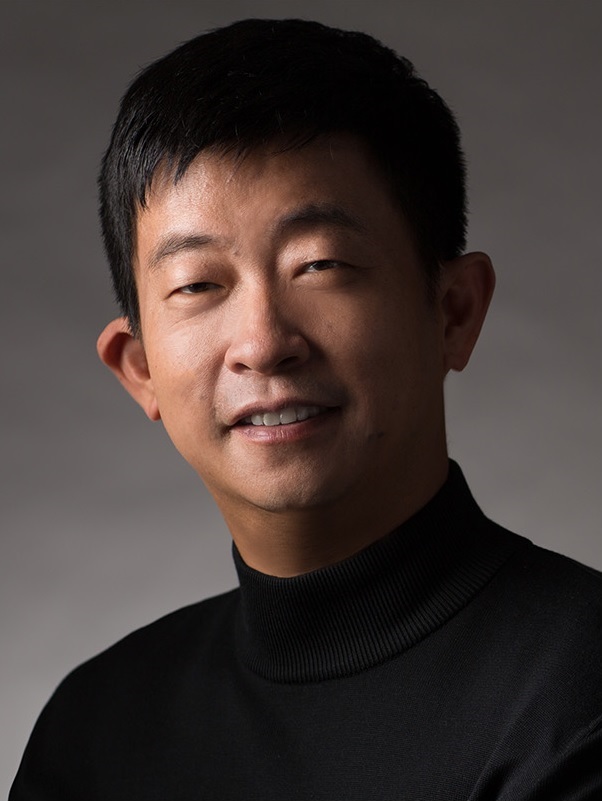}}]{Zhu Han}
		 (S’01–M’04-SM’09-F’14) received the B.S. degree in electronic engineering from Tsinghua University, in 1997, and the M.S. and Ph.D. degrees in electrical and computer engineering from the University of Maryland, College Park, in 1999 and 2003, respectively.
		
		From 2000 to 2002, he was an R\&D Engineer of JDSU, Germantown, Maryland. From 2003 to 2006, he was a Research Associate at the University of Maryland. From 2006 to 2008, he was an assistant professor at Boise State University, Idaho. Currently, he is a John and Rebecca Moores Professor in the Electrical and Computer Engineering Department as well as in the Computer Science Department at the University of Houston, Texas. Dr. Han’s main research targets on the novel game-theory related concepts critical to enabling efficient and distributive use of wireless networks with limited resources. His other research interests include wireless resource allocation and management, wireless communications and networking, quantum computing, data science, smart grid, security and privacy.  Dr. Han received an NSF Career Award in 2010, the Fred W. Ellersick Prize of the IEEE Communication Society in 2011, the EURASIP Best Paper Award for the Journal on Advances in Signal Processing in 2015, IEEE Leonard G. Abraham Prize in the field of Communications Systems (best paper award in IEEE JSAC) in 2016, and several best paper awards in IEEE conferences. Dr. Han was an IEEE Communications Society Distinguished Lecturer from 2015-2018, AAAS fellow since 2019, and ACM distinguished Member since 2019. Dr. Han is a 1\% highly cited researcher since 2017 according to Web of Science. Dr. Han is also the winner of the 2021 IEEE Kiyo Tomiyasu Award, for outstanding early to mid-career contributions to technologies holding the promise of innovative applications, with the following citation: ``for contributions to game theory and distributed management of autonomous communication networks."						
	\end{IEEEbiography}
	\vspace{-20mm}

	\begin{IEEEbiography}[{\includegraphics[width=1in,height=1.25in,clip,keepaspectratio]{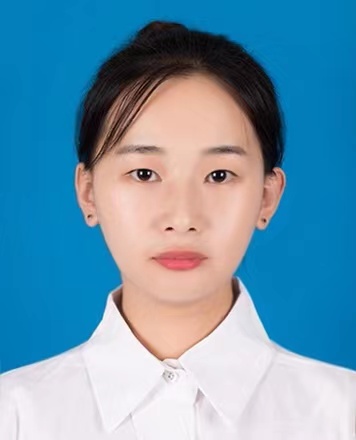}}]{Guiqi Sun}
	Guiqi Sun (Student Member, IEEE) received the M.S. degree from Qingdao University, Qingdao, China, in 2019. She is currently working toward the PH.D. degree with the State Key Laboratory of Advanced Rail Autonomous Operation, Beijing Jiaotong University, Beijing, China.
	She is a Visiting Student with Engineering Product Development, Singapore University of Technology and Design, Singapore. Her current research interests include wireless propagation channel modeling and reconfigurable intelligent surface channel modeling.					
\end{IEEEbiography}
	\vspace{-30mm}

	\begin{IEEEbiography}[{\includegraphics[width=1in,height=1.25in,clip,keepaspectratio]{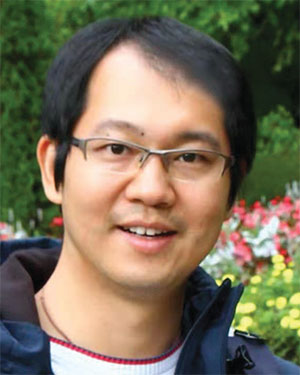}}]{Ning Wang}
		(Member, IEEE) received the B.E. degree in communication engineering from Tianjin University, Tianjin, China, in 2004, the M.A.Sc. degree in electrical engineering from The University of British Columbia, Vancouver, BC, Canada, in 2010, and the Ph.D. degree in electrical engineering from the University of Victoria, Victoria, BC, Canada, in 2013. From 2004 to 2008, he was with the China Information Technology Design and Consulting Institute, as a Mobile Communication System Engineer, specializing in planning and design of commercial mobile communication networks, network traffic analysis, and radio network optimization. From 2013 to 2015, he was a Postdoctoral Research Fellow with the Department of Electrical and Computer Engineering, The University of British Columbia. Since 2015, he has been with the School of Information Engineering, Zhengzhou University, Zhengzhou, China, where he is currently an Associate Professor. He also holds adjunct appointments with the Department of Electrical and Computer Engineering, McMaster University, Hamilton, ON, Canada, and the Department of Electrical and Computer Engineering, University of Victoria. His research interests include resource allocation and security designs of future cellular networks, channel modeling for wireless communications, statistical signal processing, and cooperative wireless communications. He was on the Technical Program Committees of international conferences, including the IEEE GLOBECOM, IEEE ICC, IEEE WCNC, and CyberC. He was on the Finalist of the Governor Generals Gold Medal for Outstanding Graduating Doctoral Student with the University of Victoria in 2013.
	\end{IEEEbiography}
	\vspace{-30mm}

	\begin{IEEEbiography}[{\includegraphics[width=1in,height=1.25in,clip,keepaspectratio]{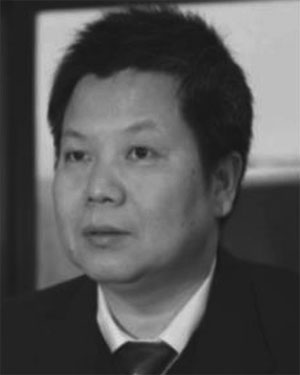}}]{Zhangdui Zhong}
		(SM’16) received the B.E. and M.S. degrees from Beijing Jiaotong University, Beijing, China, in 1983 and 1988, respectively. He is currently a Professor and an Advisor of Ph.D. candidates with Beijing Jiaotong University, where he is also currently a Chief Scientist of the State Key Laboratory of Advanced Rail Autonomous Operation. He is also the Director of the Innovative Research Team, Ministry of Education, Beijing, and a Chief Scientist of the Ministry of Railways, Beijing.
		
		He is also an Executive Council Member of the Radio Association of China, Beijing, and a Deputy Director of the Radio Association, Beijing. His interests include wireless communications for railways, control theory and techniques for railways, and GSM-R systems. His research has been widely used in railway engineering, such as the Qinghai-Xizang railway, DatongQinhuangdao Heavy Haul railway, and many high-speed railway lines in China. He has authored or co-authored seven books, five invention patents, and over 200 scientific research papers in his research area. Prof. Zhong was a recipient of the Mao YiSheng Scientific Award of China, Zhan TianYou Railway Honorary Award of China, and Top 10 Science/Technology Achievements Award of Chinese Universities.
	\end{IEEEbiography}
	
		\begin{IEEEbiography}[{\includegraphics[width=1in,height=1.25in,clip,keepaspectratio]{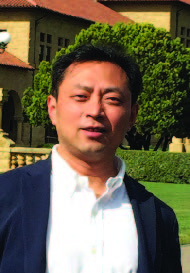}}]{Bo Ai}
		Bo Ai received the M.S. and Ph.D. degrees from Xidian University, China. He studies as a Post-Doctoral Student at Tsinghua University. He was a Visiting Professor with the Electrical Engineering Department, Stanford University, in 2015. He is currently with Beijing Jiaotong University as a Full Professor and a Ph.D. Candidate Advisor. He is the Deputy Director of the State Key Laboratory of Advanced Rail Autonomous Operation and the Deputy Director of the International Joint Research Center. He is one of the
		main people responsible for the Beijing Urban Rail Operation Control System, International Science and Technology Cooperation Base. He is also a Member, of the Innovative Engineering Based jointly granted by the Chinese Ministry of Education and the State Administration of Foreign Experts Affairs. He was honored with the Excellent Postdoctoral Research Fellow by Tsinghua University in 2007.
		
		He has authored/co-authored eight books and published over 300 academic research papers in his research area. He holds 26 invention patents. He has been the research team leader for 26 national projects. His interests include the research and applications of channel measurement and channel modeling, dedicated mobile communications for rail traffic systems. He has been notified by the Council of Canadian Academies that, based on Scopus database, he has been listed as one of the Top 1\% authors in his field all over the world. He has also been feature interviewed by the IET Electronics Letters. He has received some important scientific research prizes.
		
		Dr. Ai is a fellow of the Institution of Engineering and Technology. He is an Editorial Committee Member of the Wireless Personal Communications journal. He has received many awards, such as the Outstanding Youth Foundation from the National Natural Science Foundation of China, the Qiushi Outstanding Youth Award by the Hong Kong Qiushi Foundation, the New Century Talents by the Chinese Ministry of Education, the Zhan Tianyou
		Railway Science and Technology Award by the Chinese Ministry of Railways, and the Science and Technology New Star by the Beijing Municipal Science and Technology Commission. He was a co-chair or a session/track chair for  many international conferences. He is an IEEE VTS Beijing Chapter Vice Chair and an IEEE BTS Xi’an Chapter Chair. He is the IEEE VTS Distinguished Lecturer. He is an Editor of the IEEE TRANSACTIONS ON CONSUMER ELECTRONICS. He is the Lead Guest Editor of Special Issues of the IEEE TRANSACTIONS ON VEHICULAR TECHNOLOGY, the IEEE ANTENNAS AND WIRELESS PROPAGATION LETTERS, and the International Journal of Antennas and Propagation.
	\end{IEEEbiography}

\end{document}